\documentclass[aip,preprint]{revtex4-1}

\usepackage{graphicx}
\usepackage{amssymb}
\usepackage{amsmath}
\usepackage{color}
\usepackage[overload]{empheq}
\usepackage{setspace}

\newcommand{\Matrix}[1]{\mathsf{#1}}
\newcommand{\Vector}[1]{\mathbf{#1}}

\begin{document}

\title{Slow sound in lined flow ducts} 

\author{Y. Aur\'egan}
\email{yves.auregan@univ-lemans.fr}
\author{V. Pagneux}
\affiliation{Laboratoire d'Acoustique de l'Universit\'e du Maine,
UMR CNRS 6613
Av. O Messiaen, F-72085 LE MANS Cedex 9,
France}

\begin{abstract}
We consider the acoustic propagation in lined flow duct with a purely reactive impedance at the wall. This reacting liner has the capability to reduce the speed of sound, and thus to enhance the interaction between the acoustic propagation and the low Mach number flow ($M\simeq0.3$).  At the lower frequencies, there are typically 4 acoustic or hydrodynamic propagating modes, with 3 of them propagating in the direction of the flow. Above a critical frequency, there are only 2 propagating modes that all propagate in the direction of the flow. From the exact 2D formulation an approximate 1D model is developed to study the scattering of acoustic waves in a straight duct with varying wall impedance. This simple system, with a uniform flow and with a non-uniform liner impedance at the wall, permits to study the scattering between regions with different waves characteristics. Several situations are characterized to show the importance of negative energy waves, strong interactions between acoustic and hydrodynamic modes or asymmetric scattering.
 \end{abstract}
 
\pacs{43.28.Py, 43.50.Gf, 43.20.Mv }
\maketitle

\section{Introduction}

Acoustic liners are widely used to reduce the sound transmission in ducts with flow with applications in household appliances, ventilation systems in vehicles and buildings, IC-engines, power plants, aircraft engine.  The mitigation due to these liners is based on two principles that are generally mixed.  The first action of the liners is to dissipate acoustical energy by visco-thermal losses or by exchange of energy between the acoustical field and flow, like in the vicinity of a hole in a perforated plate with grazing flow. The second type of action is the scattering of the acoustical waves by the changes of acoustical impedance occurring for instance at the entrance and at the exit of the liners. This paper focusses on the second type of action called reacting effects and disregards the first type called dissipative effects. To do this, a waveguide with an acoustically treated wall is studied and the wall is considered as locally reacting and without dissipation. When the liner consists of cavities mounted flush to the wall (like small closed tubes in the present case), those cavities act as springs in the low frequencies limit. Then, the speed of sound is determined by the square root of the ratio between the isentropic bulk modulus (which is a measure for the “stiffness” of the fluid) and the  “mass” density. The presence of small cavities decreases the effective stiffness and, consequently, the speed of sound. The propagative acoustical waves in such systems are called "slow sound".
Recently, slow acoustic waves have attracted attention for the potential to design new acoustic devices such as metamaterials.
They have been studied both in sonic crystals\cite{cicek2012slow} and in one dimensional system\cite{Santillan2011Acoustic, Theocharis2014}. The originality of the present study is to introduce a mean flow with a velocity of the same order as the effective speed of sound. When the flow velocity is smaller than the speed of sound the regime is called "subsonic" and "supersonic" on the other case. In the subsonic regime, it will be shown that 4 modes propagate at low frequencies (wavelength much smaller than the transverse dimensions of the waveguide). Two of these modes correspond to classical acoustical waves in both direction. The two other modes do not exist without flow and are thus called HydroDynamic (HD) modes in the following.  One of these HD modes has a group velocity and a phase velocity in opposite direction. The second HD mode is a Negative Energy Wave (NEW). Globally, among the 4 modes that propagate in the subsonic regime, 3  of the modes propagate in the flow direction while one of the acoustical modes propagates against the flow. In the supersonic regime, only 2 waves  can propagate and they are in the flow direction. The problem that we consider is close to the response of fluid loaded finite plates with mean flow \cite{Crighton1991, Peake1997, arzoumanian2011stability} but it leads to a simpler analysis of interesting behaviors. 

The plan of the paper is as follows. The section \ref{Sect2} of this paper is devoted to the characterization  of the modes propagating in the low frequencies limit in a 2D duct. It will be shown that an energy flux conservation can be written in this case.
The section \ref{Sect3} describes a approximate 1D model where the effects in the transverse direction of the duct are taken into account by averaging. Albeit very simple, this 1D model has the same richness of behavior as the 2D model. In particular, as in the 2D model, an energy flux conservation is obtained and a NEW is present. In the section \ref{Sect4}, the 1D model is used to calculate the scattering proprieties of an increase or a decrease in the wall impedance. The transonic cases (from supersonic to subsonic and vice versa) are of particular interest because of the conversion of acoustical waves into HD modes. A local transonic increase of the compliance is also studied in the section \ref{Sect5} and shows an interesting propriety of total transmission in flow direction and of no transmission in the opposite direction corresponding to an "acoustical diode".

\section{Sound propagation in a 2D duct with flow and compliant wall \label{Sect2}}

\begin{figure}[h!]
\begin{center}
\includegraphics[width=0.7\columnwidth]{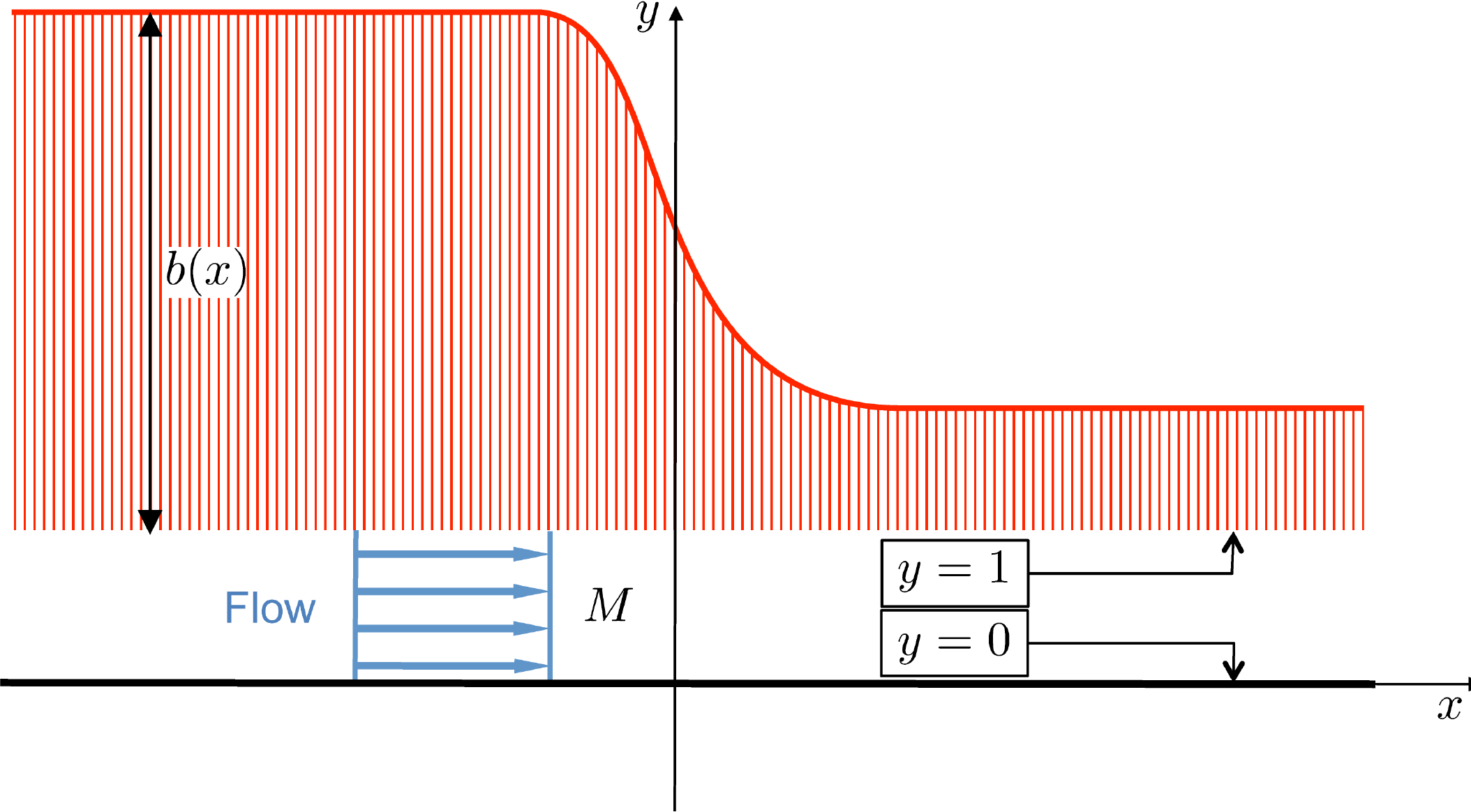}%

\begin{minipage}{10cm}
 {\small \textbf{Figure 1:} \textsl{Geometry of the problem}}
 \end{minipage}
\end{center}
\end{figure}

We consider the sound propagation in a 2D channel with a uniform flow, see Fig. 1. The lower wall is rigid. The upper wall is compliant and composed of small tubes of variable lengths. All parameters are nondimensionalized in the standard way to simplify the notation. Velocities are nondimensionalized by the speed of sound $c_0$, so that the uniform mean velocity becomes the steady flow Mach number $M$. Distances are nondimensionalized by the height of the channel $H$, time by $H/c_0$,  and pressure by $\rho_0 c_0^2$ where $\rho_0$ is the mean density.
The dimensionless equations governing the acoustic motion are then
\begin{eqnarray}
D_t \; p  &=& - \mathbf{\nabla.v} \label{eq:1} \\
D_t \; \mathbf{v} &=&  -\mathbf{\nabla} p \:, \label{eq:2}
 \end{eqnarray}
where $p$ is the pressure, $\mathbf{v}$ is the velocity and $D_t =\partial_t+M \partial_x$ is the convective derivative. 
Next, the equations are written in term of the acoustic velocity potential ($\mathbf{v}=\mathbf{\nabla} \varphi$). Eq. (\ref{eq:2}) leads to $p=-D_t \; \varphi$ and Eq. (\ref{eq:1}) leads to the classical convected wave equation:
\begin{equation}
D_t^2 \varphi -\mathbf{\nabla}^2 \varphi = 0 \,.
\label{eq:3}
\end{equation}

On the rigid wall ($y=0$), the boundary condition is  $\partial_y \varphi = 0$. 
On the compliant wall, we use the so-called "Ingard-Myers Condition" \cite{myers1980acoustic}.
This condition states that the pressure and the transverse displacement $\eta$ ($D_t\eta = v= \partial_y \varphi $) are continuous at the wall which leads to  $v = D_t (C(x) p)$ where $C(x)$ is the compliance of the wall (ratio of transverse displacement over the pressure). Hence the boundary condition at the wall $y=1$ is written: 
\begin{equation}
\partial_y \varphi = - D_t\left( C(x)  D_t \varphi \right).
\label{eq:4}
\end{equation}

The compliance of the closed tubes of length $b(x)$ at $y=1$ is given by $C(x) = \sigma \tan (b(x) \omega)/\omega$ where $\sigma$ is the percentage of open area (POA, ratio between the surface of the tubes and the total surface). In the very low frequencies limit ($\omega b \ll 1$), the compliance is simply equal to the length of the tubes $b(x)$ multiply by the POA. It means that in this limit, the closed tubes act like springs of stiffness $1/\sigma b$. To simplify the notation in the following $\sigma$ is suppose to be equal to unity (it could be integrated very easily if it differs significantly from unity) and the problem can be written globally as

\begin{align}[left=\empheqlbrace] 
\label{eq:4c}
& D_t^2 \varphi -\mathbf{\nabla}^2 \varphi = 0 \nonumber\\
&\partial_y \varphi = 0 \:\:\mathrm{at}\: y=0 \\
&\partial_y \varphi = - D_t\left( b(x)  D_t \varphi \right) \:\:\mathrm{at}\: y=1 . \nonumber
\end{align}

The impedance boundary condition with uniform flow is questionable\cite{Renou2011} and more advanced models exist \cite{Brambley2011}. The Ingard-Myers condition has been used here for simplicity. In the low frequencies limit, more complex models have been tested without qualitative changes in the behavior\cite{auregan2013}.  Furthermore, it could be noted that the low frequencies limit used here, leads to a "well posed" problem in the sense given by Brambley\cite{Brambley2009Fundamental}. 

\subsection{Dispersion equation in the 2D problem}

For uniform compliance $b$, the solution is searched under the form $\varphi=A \cosh(\alpha y) \exp (\mathrm{i}(-\omega t + k x))$ where $\alpha^2 = k^2-\Omega^2$ and $\Omega=\omega-M k$. This leads to the dispersion equation: 
\begin{equation}
\mathcal{D}(\omega,k) = \alpha \tanh(\alpha) - \frac{\tan (b \: \omega)}{\omega} \: \Omega^2 = 0
\label{eq:5}
\end{equation}
which, in the very frequency limit, becomes:
\begin{equation}
\mathcal{D}(\omega,k) = \alpha \tanh(\alpha) - b \: \Omega^2 = 0.
\label{eq:5b}
\end{equation}

Without flow in the very low frequency limit, the dispersion equation (\ref{eq:5b}) can be simplified to $k^2=(1+b)\omega^2$.  The phase velocity 
\begin{equation}
c_b = \frac{\omega}{k} = \frac{ 1}{\sqrt{1+b}}
\label{eq:5c}
\end{equation}
is always smaller than 1, meaning that $c_b$ is smaller than the speed of sound in free space. Thus the acoustic wave propagation can be significantly slowed down in a  duct with a wall which reacts locally like a spring. The phase velocity of this "slow sound" can be decreased to become of the order of the flow velocity in the duct.  In this case, dramatic effects of a flow with moderate Mach number ($M\simeq 0.3$) are expected.

\begin{figure}[h]
\begin{center}
\includegraphics[width=0.7\columnwidth]{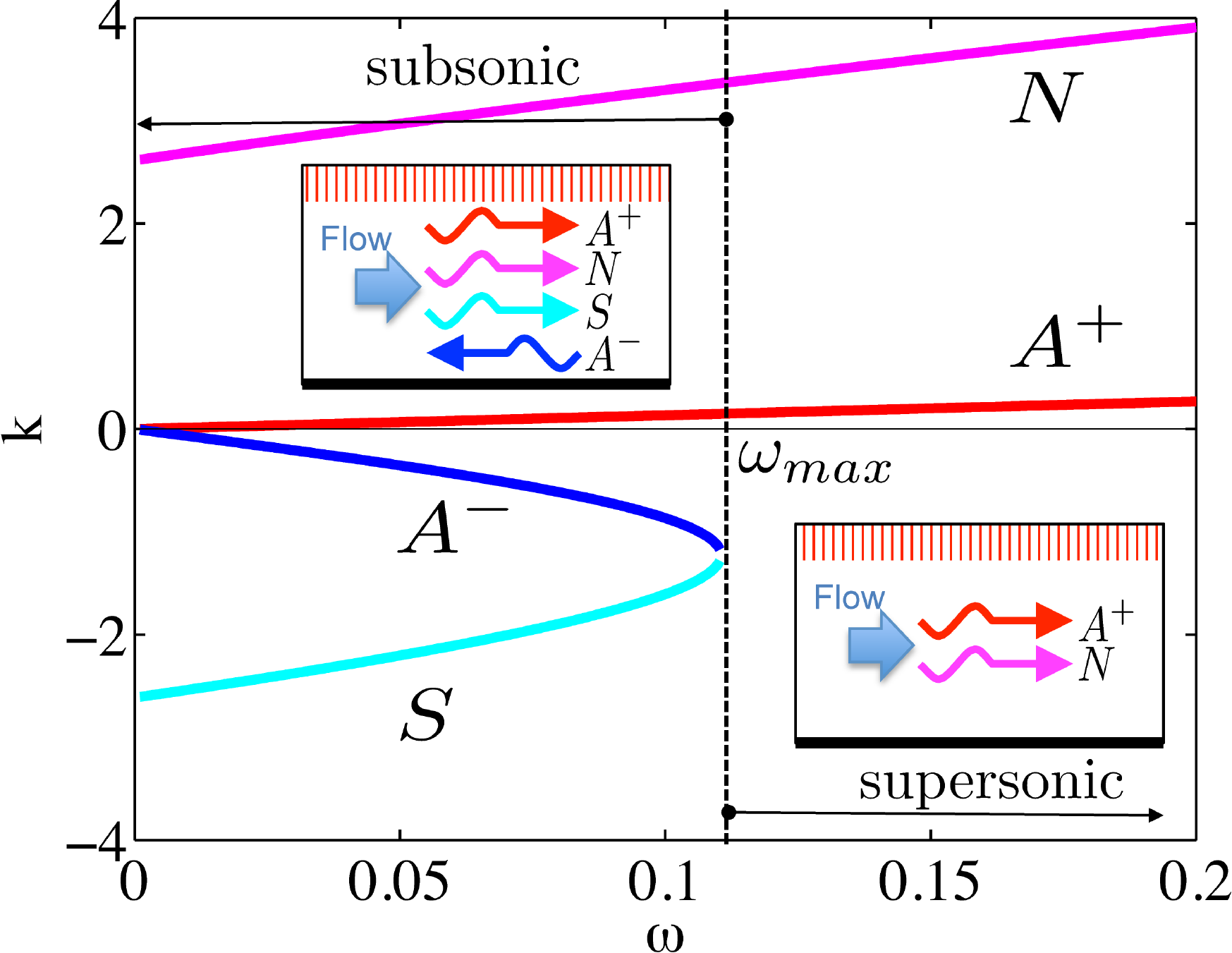}

\begin{minipage}{10cm} 
{\small \textbf{Figure 2:} \textsl{(Color online) Propagative wave numbers $k$ as a function of the frequency $\omega$ for  $b=4$ and $M=0.3$. The 2 embedded boxes give the direction of the waves propagation.}}
\end{minipage}
\end{center}
\end{figure}

The dispersion curves are displayed on Fig. 2 and show the effect of the flow. For $\omega<\omega_{max}$, Eq.  (\ref{eq:5}) has 4 real solutions corresponding to propagating modes. Two roots, labeled $S$ and $A^-$ on Fig. 2, approach to each other when  $\omega \to \omega_{max}$. They coalesce for a frequency $\omega_{max}$ above which they no longer exist as real roots i.e. as propagating waves. It is shown in Appendix \ref{AppendixI} that  when $M>c_b$ there are only two propagating modes whatever the frequency (i.e. $\omega_{max}=0$).

In the subsonic regime, i.e. when $M < c_b$ and $ \omega < \omega_{max}$, 2 of the 4 modes have a vanishing wavenumber when $\omega \to 0$. These solutions are called acoustic and, in the low frequencies limit, they propagate in both directions with the speed of sound $c_b$ corrected by the convective effects. The 2 other solutions do not exist without flow and are called HydroDynamic (HD) modes.  The solution called $S$ in Fig. 2 has a negative phase velocity $c_\Phi \equiv \omega/k$ but a positive group velocity $c_g \equiv \mathrm{d}\omega/ \mathrm{d} k$ and thus propagates in the flow direction. The last solution called $N$ has both  positive phase and group velocities and propagates in the flow direction. It can be seen from Fig. 12 in Appendix \ref{AppendixI} that $\Omega<0$ for this wave and it will be seen below that it corresponds to a Negative Energy Wave (NEW). 
In the supersonic regime, i.e. when $M>c_b$ or $\omega>\omega_{max}$, only the $A^+$ and the $N$ waves can propagate.

In summary, in the subsonic case 4 waves propagate. Three of them propagate in the flow direction ($A^+,\:S,\:N$) and one propagates against the flow ($A^-$). In the supersonic case, only 2 waves ($ A^+,\:N$) can propagate and they are in the flow direction. A NEW is always present. If $M > 1/\sqrt{1+b}$, we are always in the supersonic case. If $M < 1/\sqrt{1+b}$, the transition from subsonic to supersonic can been reach either by increasing $\omega$ at a given $b$ or by increasing $b$ at a given $\omega$. This last possibility will be used in the Section on scattering (Sec. \ref{Sect4} and \ref{Sect5}).

\subsection{Energy flux conservation of slow sound waves with flow}

Thereafter the problem is studied in the frequency domain (convention $\mathrm{e}^{-\mathrm{i} \omega t}$) where $\partial_t \equiv -\mathrm{i} \omega$ and $D_\omega = -\mathrm{i} \omega+M \partial_x$. In order to find an "energy like" equation, the Eq. (\ref{eq:3}) is classically multiplied by $\overline{\varphi}$ (the complex conjugate of $\varphi$) and is integrated on the cross section to yield: 
\begin{eqnarray}
& & \Im \mathrm{m} \left( \int_0^1 \overline{\varphi}( \partial^2_x \varphi + \partial^2_y \varphi - D^2_\omega  \varphi )\:\mathrm{d}y \right) = \nonumber \\
& & \Im \mathrm{m} \left( \partial_x \left(\int_0^1 (\overline{\varphi} \partial_x \varphi - M \overline{\varphi} D_\omega  \varphi )\:\mathrm{d}y \right) + 
\left[ \overline{\varphi} \partial_y \varphi \right]_0^1  \right) = \nonumber \\
& & \partial_x \bigg( \Im \mathrm{m} \Big( \int_0^1 (\overline{\varphi} \partial_x \varphi  - M \overline{\varphi} D_\omega  \varphi )\:\mathrm{d}y  \\
       & & - M \overline{\varphi}(x,1) b(x) D_\omega \varphi(x,1)   \Big) \bigg) = 0 ,\nonumber 
\label{eq:6}
\end{eqnarray}
where the relation
\begin{equation}
\Im \mathrm{m} \left( \overline{g} D_\omega (f(x) D_\omega g) \right) = \partial_x \left( \Im \mathrm{m} \left( M \overline{g} f(x) D_\omega g \right)\right) 
\label{eq:7}
\end{equation}
valid for any function $g$ and any real function $f$ had been used. Thus the quantity 

\begin{eqnarray}
J &=& \Im \mathrm{m} \bigg( \int_0^1   \overline{\varphi} (\partial_x \varphi - M  D_\omega  \varphi )\:\mathrm{d}y \nonumber \\
& &  - M \overline{\varphi}(x,1) b(x) D_\omega \varphi(x,1)   \bigg)
\label{eq:EC}
\end{eqnarray}
is conserved along $x$. This expression is identical to the expression of the energy flux proposed by M\"ohring\cite{Mohring2001Energy}.
This energy flux can be computed for each mode $m$ (normalized by its value at $y=1$):  $\varphi_m = \cosh(\alpha_m y) \mathrm{e}^{\mathrm{i} k_m x} /\cosh(\alpha_m)$ where $\alpha_m$ is one of the solutions of the dispersion equation, $\alpha_m^2 =k_m^2-\Omega_m^2$ and $\Omega_m= \omega-M k_m$ :
\begin{equation*}
J_m =    \left( (k_m + M \Omega_m) \frac{\sinh (2 \alpha_m)+2\alpha_m }{4\alpha_m \cosh^2 (\alpha_m)} + M b \, \Omega_m  \right).
\end{equation*}

The value of $J_m$ is displayed on Fig. 3 for the 4 propagating modes. The modes $A^+$ and $S$, that propagate in the flow direction, have positive energy fluxes. The modes $A^-$, that propagates against the flow, has negative energy flux. The mode $N$, that propagates in the flow direction, has a negative energy flux. It means that this mode is a NEW\cite{Crighton1991}. This last property will have important consequences in the results presented afterwards.

\begin{figure}[h]
\begin{center}
\includegraphics[width=0.7\columnwidth]{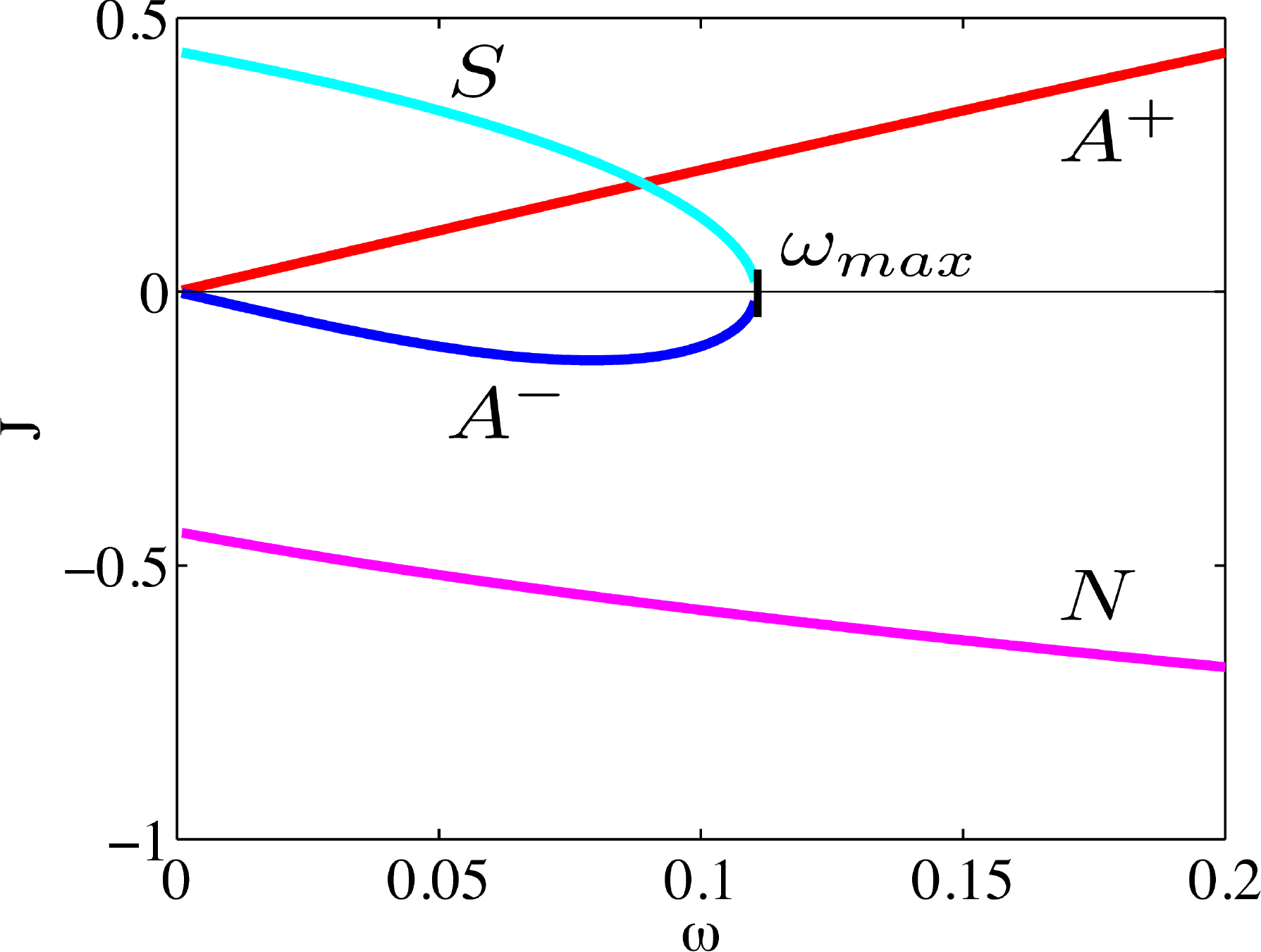}%

\begin{minipage}{10cm}
 {\small \textbf{Figure 3:} \textsl{(Color online) Energy flux $J$ as a function of the frequency $\omega$ for  $b=4$ and $M=0.3$.}}
 \end{minipage}
\end{center}
\end{figure}

\section{1D model \label{Sect3}}

\subsection{1D approximation}

In order to simplify the analysis of the problem, we are looking for a 1D model that conserved the main proprieties of the 2D problem: The dispersion relation has to give the same number of propagating modes as the 2D model and a conserved energy flux has to be defined. For that, we integrate the 2D equation (\ref{eq:3}) along $y$ and we get the exact expression:
\begin{equation}
D^2_t  \left(\int_0^1\varphi \:\mathrm{d}y \right) - \partial^2_x \left(\int_0^1\varphi \:\mathrm{d}y \right)  - \partial_y \varphi(x,1)=0
\label{eq:1D1}
\end{equation}
which is associated to the boundary conditions in (\ref{eq:4c}).
A simplification can be achieved if we now assume that the $y$ derivative of $\varphi$ at the compliant wall can be written:
\begin{equation}
\partial_y \varphi(x,1) = a_1 V (x) + a_2 F (x)
\label{eq:1D2}
\end{equation}
where $a_1$ and $a_2$ are two real constants and where $V$ and $F$ are defined as the two functions appearing in (\ref{eq:1D1}) and in the boundary condition at $y=1$ in Eq. (\ref{eq:4c}):
\begin{equation*}
V(x) = \varphi(x,1) \:\:\: \mathrm{and} \:\:\: F(x)= \int_0^1\varphi \:\mathrm{d}y.
\end{equation*}
This leads to the system of two coupled ODEs
\begin{align}[left=\empheqlbrace] 
 D^2_t F- \partial^2_x F&= a_1 V  + a_2 F \label{eq:1D3}\\
D_{t} (\,b \; D_{t} V )&= -(a_1 V + a_2 F) \label{eq:1D4}
\end{align}

The real constants $a_1$ and $a_2$  can be chosen freely. For instance, for a parabolic approximation such as $\varphi = C_1+C_2 y^2$, the constants are $a_1=-a_2=3$.

\subsection{Dispersion relation}

When $b$ is constant, looking for a solution under the form $ \exp (\mathrm{i}(-\omega t + k x))$, leads to the dispersion equation expressed as a function of the frequency in the moving frame $\Omega= \omega-Mk$:

\begin{equation}
\left( \Omega^2-k^2+a_2 \right)\left( b \, \Omega^2 - a_1  \right) + a_1 a_2=0
\label{eq:1D5}
\end{equation}

The solutions in term of $\omega$ versus $k$ are plot on Fig. 4 and compared to the results of the 2D model. The agreement between the two model is good when the coefficient $a_1=-a_2$ is chosen in such a way that the value of $k$ when $\omega \to 0$ of the modes $N$ and $S$ are closed in 1D and 2D
model. When $a_1=-a_2$, the wavenumbers for $\omega \to 0$ are:
\begin{equation}
k_{A^\pm} = \frac{\pm \sqrt{1+b}\:\: \omega}{1 \pm M \sqrt{1+b}}  \:\:\: \mathrm{and} \:\:\:
k_{N,S} =\pm \sqrt{\frac{a_1(1-M^2(1+b))}{b M^2 (1-M^2)}}
\end{equation}

\begin{figure}[h!]
\begin{center}
\includegraphics[width=0.7\columnwidth]{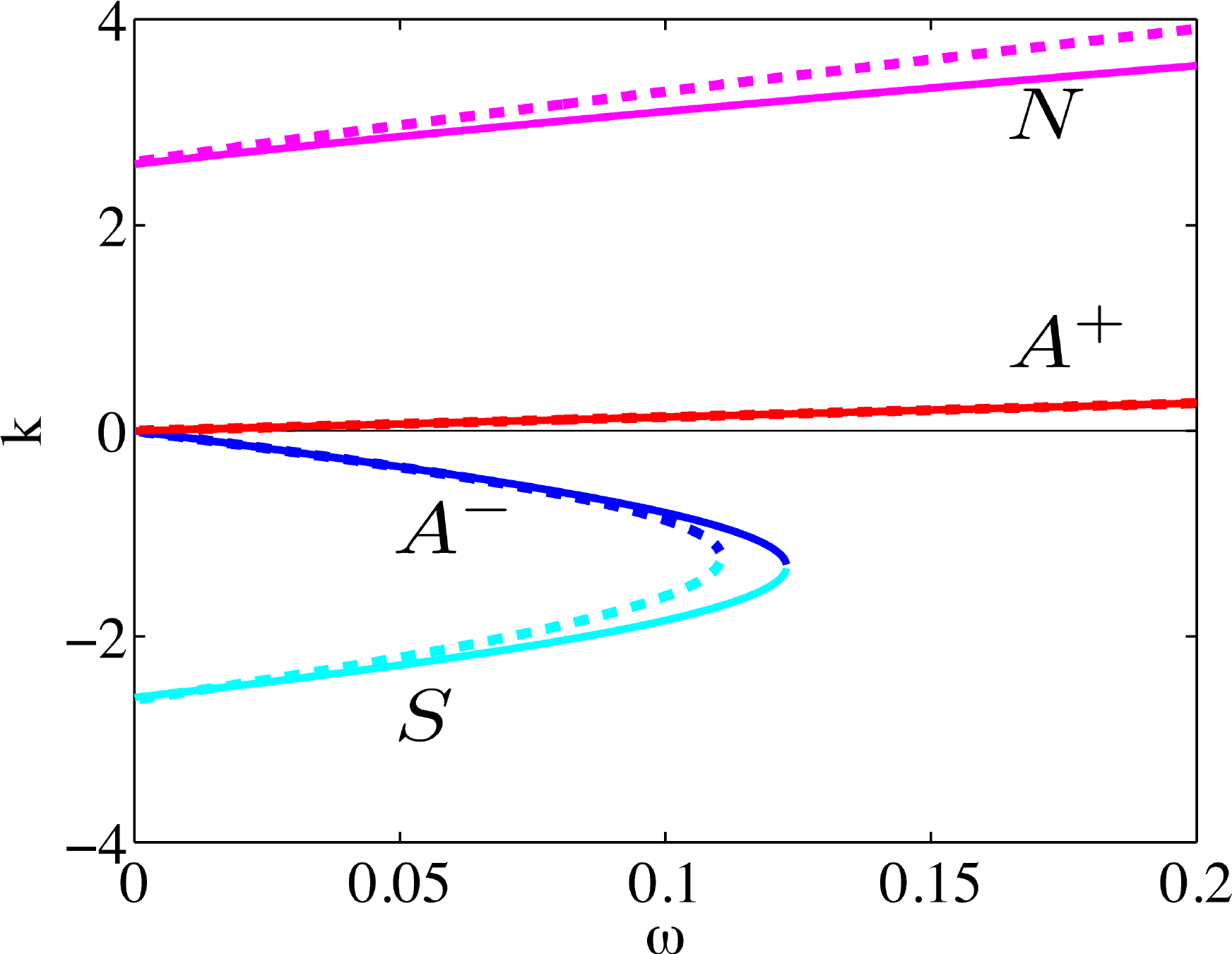}%

\begin{minipage}{10cm}
 {\small \textbf{Figure 4:} \textsl{(Color online) Propagative wave numbers $k$ as a function of the frequency $\omega$ for  $a_1=-a_2=4$, $b=4$ and $M=0.3$. The solid lines represent the solutions of the 1D model while the dashed lines represent the solution of the 2D model (see Fig. 2).}}
 \end{minipage}
\end{center}
\end{figure}

\subsection{Evolution equations}

A set of first order evolution equations can be derived from the Eqs.  (\ref{eq:1D3}) and (\ref{eq:1D4}) by introducing  $G$ and $W$ with  $G \equiv (\omega + \mathrm{i} (1-M) \partial_x) F$ and $(\omega + \mathrm{i} M\partial_x) W \equiv -(a_1 V + a_2 F)$. In vectorial notation, the evolution equation is:
\begin{equation}
-\mathrm{i}\partial_x \Vector{X} = \Matrix{Q} \Vector{X}
\:\:\:\:\text{where}\:\:\:\:
\Vector{X} = \begin{pmatrix}F \\ G \\ V\\W \end{pmatrix}
\label{Eq_evol}
\end{equation}
and 
\begin{equation*}
\Matrix{Q}=  \begin{bmatrix} 
 -\omega/(1-M)& 1/(1-M)           &     0              &0\\ 
 a_2/(1+M)      & \omega/(1+M) & a_1/(1+M)   &0\\ 
 0                    & 0                     &\omega/M      &1/(Mb)\\
a_2/M             & 0                     &a_1/M            &\omega/M\\
 \end{bmatrix}
\end{equation*}

The eigenvalues of the matrix $\Matrix{Q}$ are the four $k_m$ solutions of the dispersion equation (\ref{eq:1D5}) and the eigenvectors $\Vector{X}_m$ give a relation for each modes between the mean value of the velocity potential over the section $F_m$ and its value at the wall $V_m$. Note that, at low frequencies, the $A^+$ and $A^-$ modes are quasi plane while the $S$ and $N$ are more localized along the compliant wall. 

\subsection{Energy flux conservation in the 1D model}

To obtain an energy flux conservation, we multiply the Eq. (\ref{eq:1D3}) by $\overline{F}$ and Eq. (\ref{eq:1D4}) by $\overline{V}$ and we make use of the relation (\ref{eq:7})
:
\begin{eqnarray}
\partial_x \left( \Im \mathrm{m} \left( \overline{F}\partial_x F - M \overline{F}D_\omega F\right) \right) &=& -a_1 \Im \mathrm{m} \left(\overline{F}V  \right)  \label{eq:1D8a}\\
\partial_x \left( \Im \mathrm{m} \left(M \, b \overline{V} D_\omega V  \right) \right)&=& -a_2 \Im \mathrm{m} \left(\overline{V}F\right)  \label{eq:1D8b}
\end{eqnarray}
Thus the quantity
\begin{equation}
I = \Im \mathrm{m} \left( \overline{F}(\partial_x F - M D_\omega F) + \frac{a_1}{a_2} M \, b \overline{V} D_\omega V   \right) 
\label{eq:1DEC}
\end{equation}
is conserved along $x$. It can be noticed that if $a_1 = -a_2$, the 1D energy flux conservation (\ref{eq:1DEC}) becomes very similar to the exact energy flux conservation in 2D, see equation (\ref{eq:EC}). In this case, the energy flux of any mode ($m= A^+,\:A^-,\:S,\:N$) is given by:
\begin{equation}
I_m =  (k_m (1-M^2) + M \omega ) |F_m|^2 +   M \, b (\omega-k_m M) |V_m|^2.
\label{eq:1DECm}
\end{equation}
As in the 2D case, the modes $A^+$ and $S$ have positive energy fluxes while the modes $A^-$ and $N$ have negative energy flux. The mode $N$, propagating to the right, is thus a NEW, as in the 2D model.

The 1D model reproduces correctly all the main physical ingredients (dispersion and energy flux conservation) that are present in the 2D model. This model will be used in the next section to study the scattering induced by a change in the wall compliance in the next section.

\section{Scattering by a change in the wall compliance \label{Sect4}}

\begin{figure}[h!]
\begin{center}
\includegraphics[width=0.7\columnwidth]{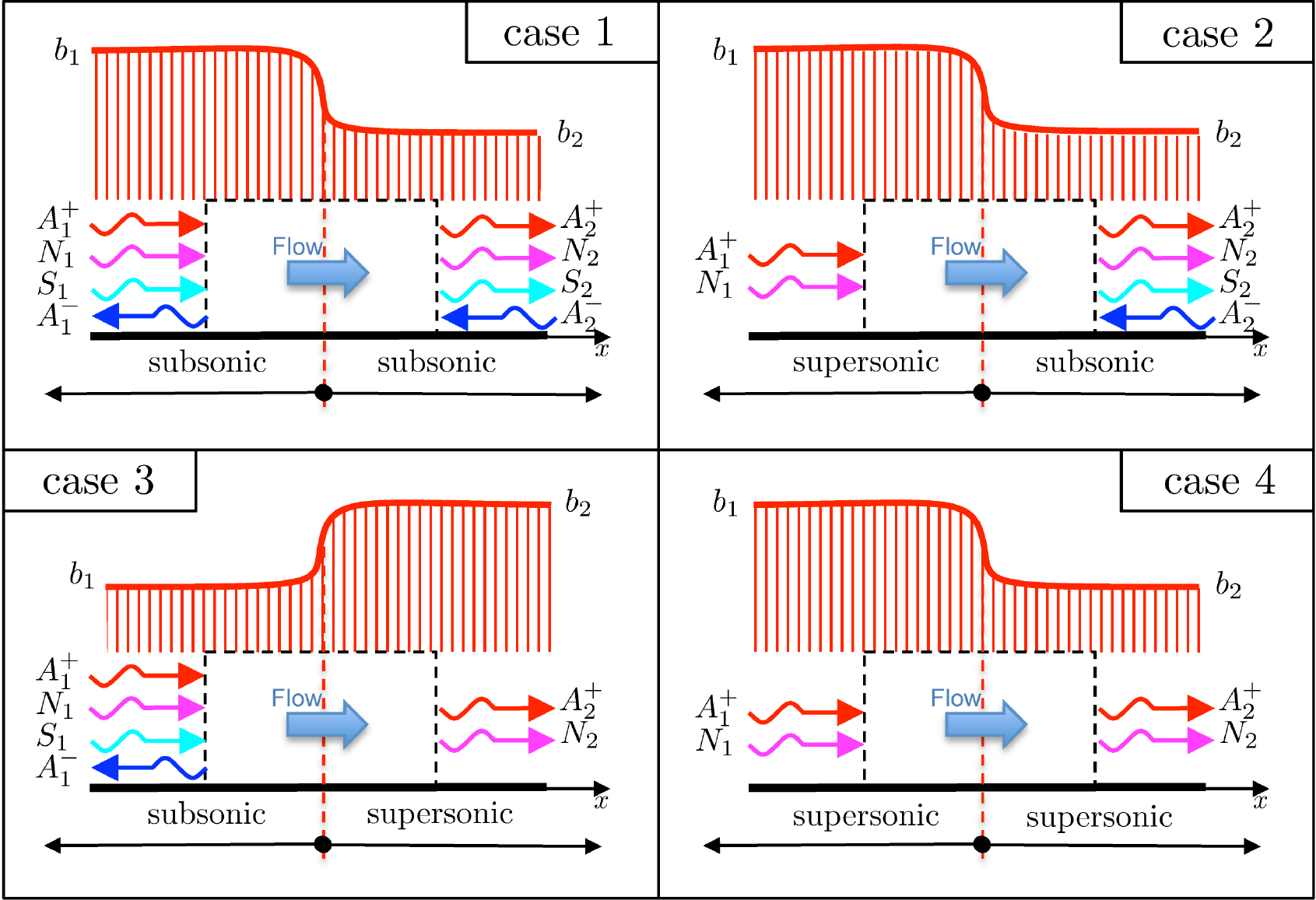}%

\begin{minipage}{10cm}
 {\small \textbf{Figure 5:} \textsl{(Color online) Scattering problems by a change in the wall compliance. The  propagative waves are given for the 4 cases considered. Case 1: subsonic everywhere (The scattering matrix $\Matrix{S}$ is $4 \times 4$). Case 2: transcritical variation of the compliance from supersonic to subsonic ($\Matrix{S}$ is $3 \times 3$). Case 2: transcritical variation of the compliance from subsonic to supersonic ($\Matrix{S}$ is $3 \times 3$). Case 4: supersonic everywhere ($\Matrix{S}$ is $2 \times 2$).}}
 \end{minipage}
\end{center}
\end{figure}

We consider the problem defined in Fig. 5: The compliance of the wall is changing around $x=0$ from the value $b_1$  ($x<0$) to a value $b_2$ ($x>0$), the flow being in the positive $x$-direction. The 4 cases indicated on Fig. 5 will be considered.

\subsection{Subsonic case (case 1) }

In the case 1, the problem is subsonic everywhere. Upstream, at left, there are 3 incoming and 1 outgoing  waves. Downstream, at right, there are 1 incoming and  3 outgoing waves.
The upstream and downstream propagative field can be described by: 
\begin{eqnarray}
\Vector{X}_j(x) &=& a_j^+  \hat{\Vector{X}}^{A^+}_j \: \mathrm{e}^{\mathrm{i} k^{A^+}_j x}
+ n_j  \hat{\Vector{X}}^{N}_{j} \mathrm{e}^{\mathrm{i} k^{N}_{j} x} \nonumber \\
& & + s_j  \hat{\Vector{X}}^{S}_{j} \mathrm{e}^{\mathrm{i} k^{S}_{j} x}
+a_j^-  \hat{\Vector{X}}^{A^-}_{j} \: \mathrm{e}^{\mathrm{i} k^{A^-}_{j} x}
\label{eq:field}
\end{eqnarray}
where $j=1$ or $2$ labels the region, the hat indicates that the modes have been normalized such that their energy flux is $1$ for the modes $A^+$ and $S$ and $-1$ for the modes $A^-$ and $N$ ($ \hat{\Vector{X}}^m = \Vector{X}^m/ \sqrt{|I_m|}$). 

The effect of the compliance variation is described by the scattering matrix linking the 4 outgoing waves $\Vector{B}$ to the 4 incoming waves $\Vector{A}$:
\begin{equation}
\Vector{B} =  \Matrix{S} \Vector{A}
\:\: \mathrm{where} \: \Vector{B}= \begin{pmatrix}a_2^+ \\ n_2 \\ s_2 \\a^-_1\end{pmatrix}
\:\:  \mathrm{and} \: \Vector{A}= \begin{pmatrix}a_1^+ \\ n_1\\ s_1  \\ a^-_2 \end{pmatrix}
\label{eq:Scatt}
\end{equation}
 where the coefficients of $\Matrix{S}$ are classically given by $S_{m\,n}$ with $n$ the incident mode and $m$ the outgoing mode.
 
 The classical unitary relation \cite{Pagneux2004} for a conservative system is replaced in our case (see Appendix \ref{AppendixII}) by
 
 \begin{equation}
 \overline{ \Matrix{S}}^T \Matrix{J} \Matrix{S}=\Matrix{J}
  \label{eq:MatConserv}
\end{equation}
where $\Matrix{J} = \mathrm{diag}(1,\, -1,\, 1,\, 1)$
and the superscript $T$ denotes the transpose operation.

Even if the scattering matrix can be easily computed for a continuous variation of $b(x)$ by numerical integration, for the sake of simplicity only the results for discontinuous variation of $b$ at $x=0$ will be presented. It can be seen from Eq. (\ref{Eq_evol}) that the functions $F$, $G$, $V$ and $W$ are continuous when $b$ is discontinuous while the slope of $V$ is discontinuous at $x=0$. The continuity of $ \Vector{X}$ at $x=0$ can be written in vectorial form, separating the incoming and the outgoing waves: 
\begin{eqnarray}
& &\underbrace{\left [  \hat{\Vector{X}}^{A^+}_{2},  \hat{\Vector{X}}^{N}_{2},  \hat{\Vector{X}}^{S}_{2}, - \hat{\Vector{X}}^{A^-}_{1}\right ]}_{\textstyle \Matrix{V_O}}
\begin{pmatrix}a_2^+ \\ n_2\\ s_2  \\ a^-_1 \end{pmatrix} = \nonumber \\
& &\underbrace{\left [  \hat{\Vector{X}}^{A^+}_{1},  \hat{\Vector{X}}^{N}_{1},  \hat{\Vector{X}}^{S}_{1}, -  \hat{\Vector{X}}^{A^-}_{2}\right ]}_{\textstyle \Matrix{V_I}}
\begin{pmatrix}a_1^+ \\ n_1\\ s_1  \\ a^-_2 \end{pmatrix} .
\end{eqnarray}
The scattering matrix is then computed by:
\begin{equation}
\Matrix{S} = \Matrix{V_O^{-1}} \: \Matrix{V_I}
\end{equation}
As an example of the scattering matrix elements, the value of the outgoing waves when the wave $A_1^+$ is incident are plotted in Fig. 6.

\begin{figure}[h!]
\begin{center}
\includegraphics[width=0.7\columnwidth]{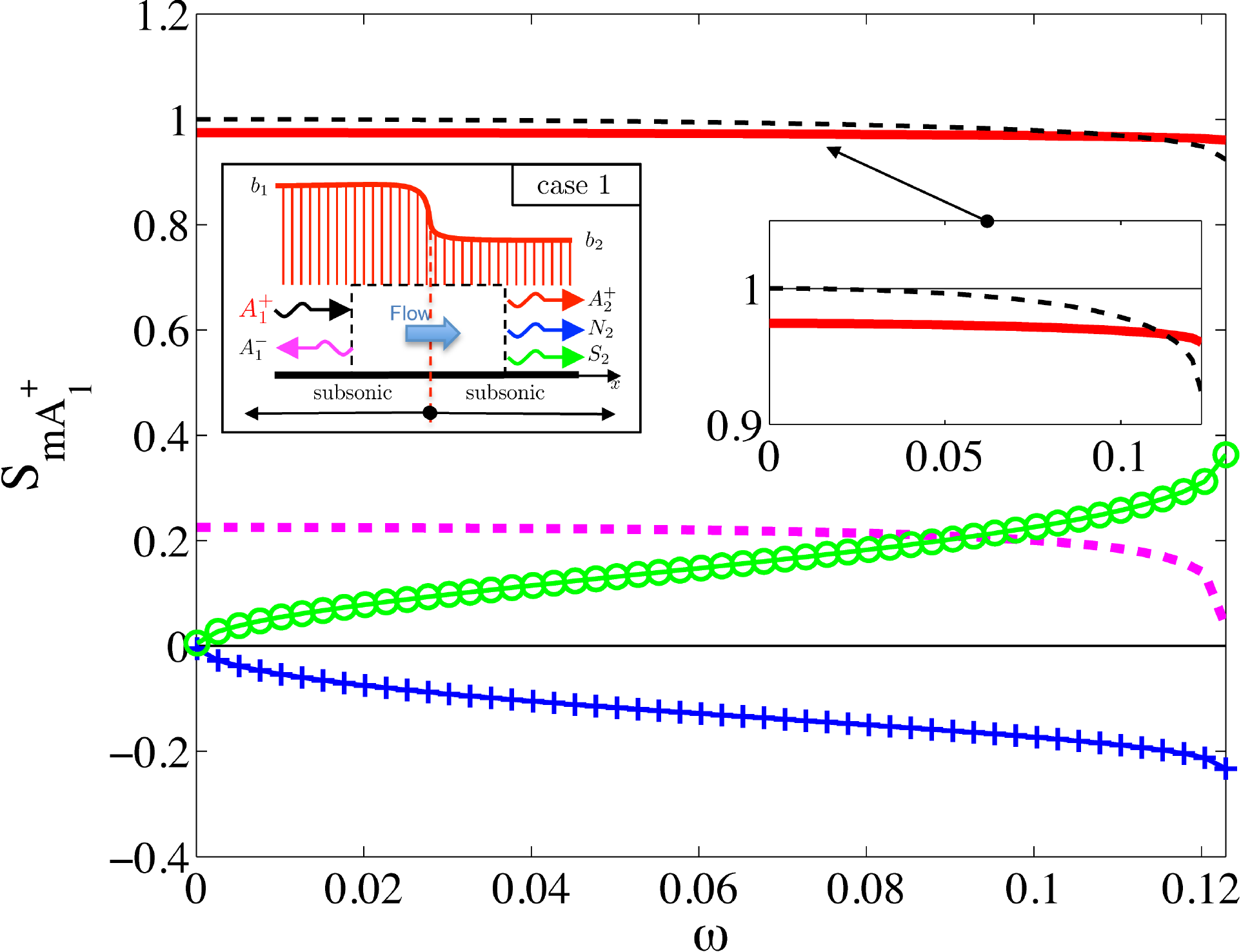}%

\begin{minipage}{10cm}
 {\small \textbf{Figure 6:} \textsl{(Color online) Value of the outgoing waves when the wave $A_1^+$ is incident ($a_1^+=1$) for $M=0.3$, $b_1=4$, $b_2=1$ and $a_1=-a_2=4$. The 4 curves represent the 4 outgoing waves: 
\textcolor{red}{---}: $A^+_2$, 
\textcolor{blue}{+++}: $N_2$, 
\textcolor{green}{ooo}: $S_2$ and 
\textcolor{magenta}{- - -}: $A^-_1$.
The thin dashed line represent the value of $| T_{A}^+|^2+| R_{A}^+|^2 $. The embedded figure is a zoom around 1. }}
 \end{minipage}
\end{center}
\end{figure}

It can be seen that the wave $A_1^+$ is mainly transmitted on $A_2^+$ and some acoustical reflection on $A^-_1$ occurs. The acoustical transmission and reflection are nearly constant up to the value $\omega_{max} = 0.1228$ where the propagation becomes sonic in the tube with the larger $b$. We can define the acoustical transmission coefficients by $ T_{A}^+=S_{a_2^+a_1^+}$ and the acoustical reflection coefficients by $ R_{A}^+=S_{a_1^-a_1^+}$. It can be seen from Fig. 6 that an energy-like conservation for the acoustical waves can be written:  $| T_{A}^+|^2+| R_{A}^+|^2$ is close and always smaller than 1. There is some conversion from the acoustical modes to the HD modes $N$ and $S$. This conversion increases linearly from 0 at $\omega=0$ but those modes are such that their energies are opposite to fulfill  the exact energy conservation, from Eq. (\ref{eq:1DEC}):
\begin{equation*}
| T_{A}^+|^2 + | R_{A}^+|^2+ |S_{s_2a_1^+} |^2- |S_{n_2a_1^+}|^2=1.
\end{equation*}
When HD modes $N_1$ or $S_1$ are incident (not displayed), they are mainly transmitted on the same mode $N_2$ resp. $S_2$ and some extra transmission occurs on $S_2$ resp. $N_2$. The conversion to acoustical modes is small. 

The overall picture of the subsonic case is that both the acoustical modes and the HD modes are rather independent. Some small conversions exist between those two families of modes. When the problem is near transonic,  the coupling between the different kind of modes becomes larger. 

\subsection{Transonic case (case 2 ) }

In the transonic case 2, on the upstream side $M> 1/ \sqrt{1+b_1}$ and the propagation is supersonic whatever $\omega$. In this case, the modes ${S}_{1}$ and $A^-_1$ are no longer propagative but they are transformed into 2 evanescent modes that are complex conjugate: $E^+$ and $E^-$. The $E^+$ mode is defined such as it decreases when $x$ increases ($\Im \mathrm{m}(k_{E^+}) > 0$). In this transonic case, 2 incoming waves are present upstream and 1 incoming and 3 outgoing waves are present downstream. 
Therefore the scattering matrix $\Matrix{S}$ is now a 3 $\times$ 3 matrix. To apply the continuity of $\Vector{X}$ at $x=0$, it is necessary to take into account the evanescent mode that decays in the $-x$ direction ($E^-$). The output matrix $\Matrix{V_O}$ is transformed into 
$\Matrix{V_O} = [  \hat{\Vector{X}}^{A^+}_{2},  \hat{\Vector{X}}^{N}_{2},  \hat{\Vector{X}}^{S}_{2}, - \hat{\Vector{X}}^{E^-}_{1} ]$
while the input matrix $\Matrix{V_I}$ is reduced to 
$\Matrix{V_I} =  [  \hat{\Vector{X}}^{A^+}_{1},  \hat{\Vector{X}}^{N}_{1},  -  \hat{\Vector{X}}^{A^-}_{2} ]$. The scattering matrix is obtained from $\Matrix{V_O^{-1}} \: \Matrix{V_I}$ by removing the last line linked to the evanescent mode. The coefficient of the scattering matrix are now complex numbers and the absolute values of 3 of these coefficients (when $A_2^-$ is incident) are shown in Fig. 7.

\begin{figure}[h!]
\begin{center}
\includegraphics[width=0.7\columnwidth]{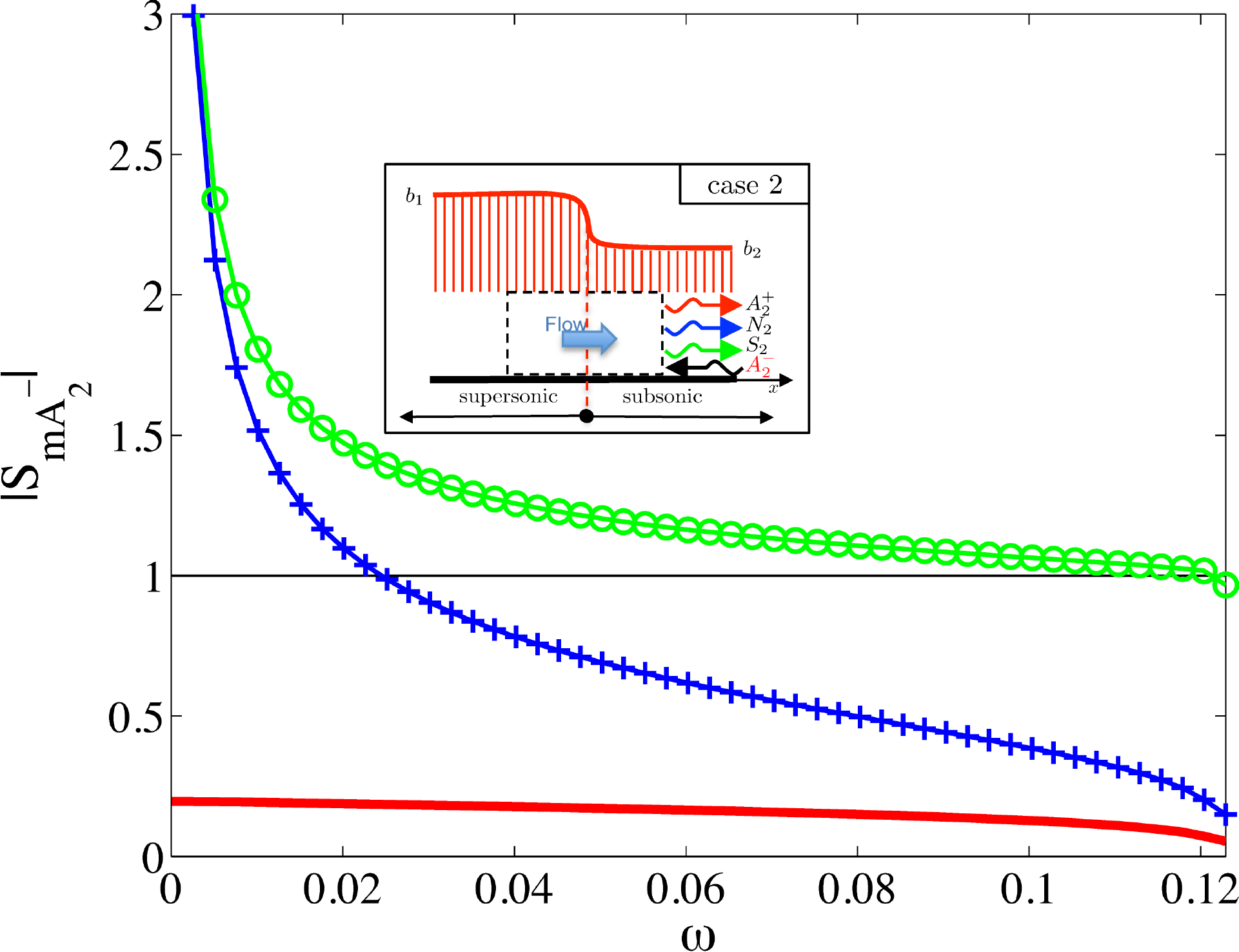}%

\begin{minipage}{10cm}
 {\small \textbf{Figure 7:} \textsl{(Color online) Absolute value of the outgoing waves when the wave $A_2^-$ is incident ($a_2^-=1$) for $M=0.3$, $b_1=12$, $b_2=4$ and $a_1=-a_2=4$.
The 3 curves are linked to the 3 outgoing waves:
\textcolor{red}{---}: $A^+_2$, 
\textcolor{blue}{+++}: $N_2$ and  
\textcolor{green}{ooo}: $S_2$. }}
 \end{minipage}
\end{center}
\end{figure}

The first striking point on the Fig. 7, is the divergence of 2 of the curves at $\omega \to 0$. When $\omega \to 0$, the energy flux, see Eq. (\ref{eq:1DECm}), for all the acoustical modes go linearly to 0 while the energy of HD modes (with a non-zero value of $k$ at $\omega = 0$) do not go to zero in the subsonic region. In the supersonic region, the energy of the $N$ mode goes to 0 when $\omega \to 0$. To ensure continuity in $\Vector{X}$, the amplitude of some of the mode coefficients has to go to infinity (like $\omega^{-1/2}$) while the amplitudes of the eigenvectors go to 0 due to the normalization. 

When an acoustical wave is incident from the upstream side, its transmission is closed to 1 (not displayed). Nevertheless, two HD modes with opposite energy flux are created. When the mode $N$ is incident upstream, this wave is mainly transmitted with an amplitude larger than 1 due to the negative energy characteristic of the wave. The $S$ wave is also created but the conversion into acoustical wave is weak. 
Interestingly, when an acoustical wave is send downstream $A^-_2$, see Fig. 7, it is mainly converted into $S$ and $N$ waves and the reflection on the acoustical wave $A_2^+$ is weak (the absolute value of the acoustical reflection coefficient is of the order of 0.15). Thus, most of the incident acoustical energy had been transferred to the HD modes. This fact is also illustrated in Fig. 8  where a temporal simulation of the Eqs. (\ref{eq:1D3}) is given. It can be also remarked in this figure that the group velocity of the 2 HD modes are close (they are equal when $\omega \to 0$)  and much smaller than the group velocity of $A^+_1$ (resp. 0.170, 0.185 and 0.747 in the present case).

\begin{figure}[h!]
\begin{center}
\includegraphics[width=0.7\columnwidth]{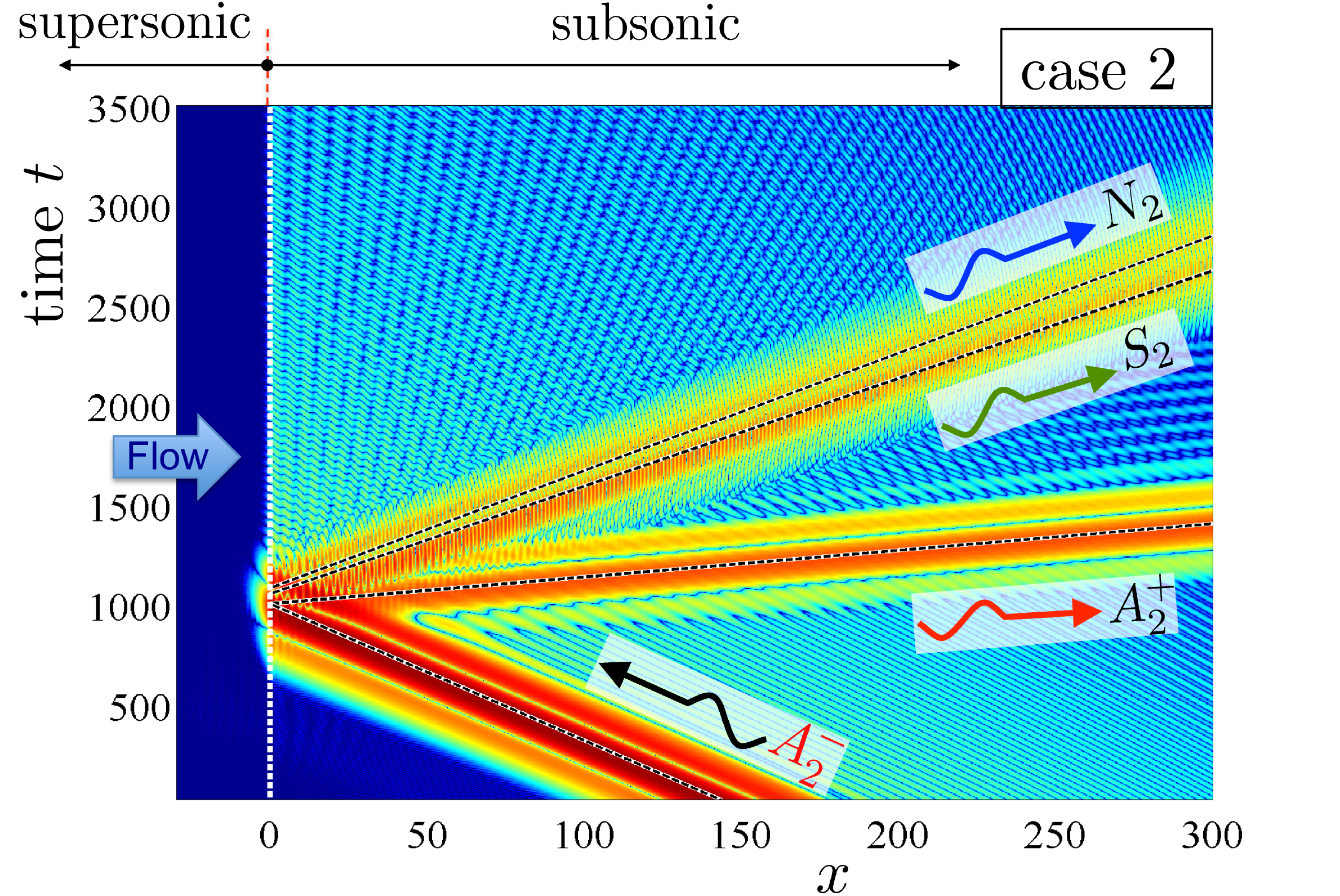}%

\begin{minipage}{10cm}
 {\small \textbf{Figure 8:} \textsl{(Color online) Temporal simulation of the variable $F$ in the case 2. The different colors represent the absolute value of $F$ in logarithmic scale for $M=0.3$, $b_1=12$, $b_2=4$ and $a_1=-a_2=4$. An $A^-_2$ pulse is send at $t=0$ with a central frequency $\omega_s = 0.015$. The slope of the black dashed lines is determined by calculating the group velocity of each modes. }}
 \end{minipage}
\end{center}
\end{figure}

The region $x<0$ is a region which cannot be excited from the outside, although waves can escape from it. In particular, NEW can escape from this region. In this sense, it can be seen as an acoustical analogous of a white hole in general relativity \cite{barcelo2011analogue}.

\subsection{Transonic case (case 3 ) }

The transonic case 3, can be treated with a  method similar to the case 2 except that the evanescent wave that had to be taken into account is $E^+_2$. The 
output matrix $\Matrix{V_O}$ is transformed into 
$\Matrix{V_O} = [ \hat{\Vector{X}}^{A^+}_{2},  \hat{\Vector{X}}^{N}_{2},   \hat{\Vector{X}}^{E^+}_{2}, -  \hat{\Vector{X}}^{A^-}_{1} ]$
while the input matrix $\Matrix{V_I}$ is reduced to 
$\Matrix{V_I} =  [  \hat{\Vector{X}}^{A^+}_{1},  \hat{\Vector{X}}^{N}_{1},  -  \hat{\Vector{X}}^{S}_{1} ]$. The scattering matrix is obtained from $\Matrix{V_O^{-1}} \: \Matrix{V_I}$ by removing the 3th line linked to the evanescent mode. 

\begin{figure}[h!]
\begin{center}
\includegraphics[width=0.7\columnwidth]{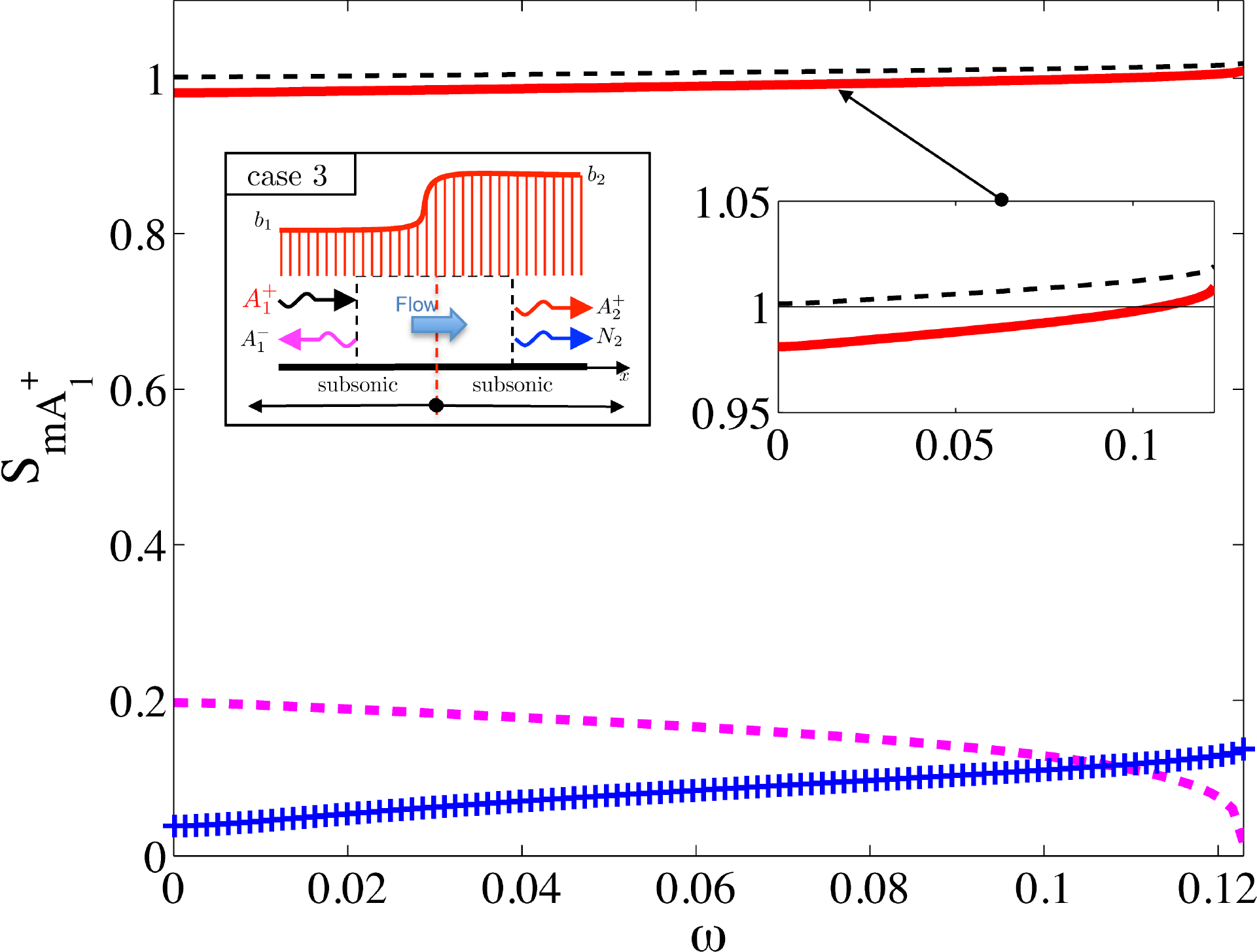}%

\begin{minipage}{10cm}
 {\small \textbf{Figure 9:} \textsl{(Color online) Absolute value of the outgoing waves when the wave $A_1^+$ is incident ($a_1^+=1$)  for $M=0.3$, $b_1=4$, $b_2=12$ and $a_1=-a_2=4$.
The 3 curves are linked to the 3 outgoing waves. 
\textcolor{red}{---}: $A^+_2$, 
\textcolor{blue}{+++}: $N_2$ and  
\textcolor{magenta}{- - -}: $A^-_1$.
The thin dashed line represent the value of $| T_{A}^+|^2+| R_{A}^+|^2 $. The embedded figure is a zoom around 1.  }}
 \end{minipage}
\end{center}
\end{figure}

When an acoustical mode is incident upstream, see Fig. 9, its transmission is again close to 1 with a small acoustical reflection. As a $N$ wave is created, the acoustical energy increases and $| T_{A}^+|^2+| R_{A}^+|^2$ is close and always greater than 1. When an HD modes $N$ or $S$ is incident (not displayed), it creates a transmitted $N$ mode and a large part is reflected as an acoustical mode. Only a small part is transmitted as an acoustical mode.

The region $x>0$ is a region from which no wave can escape.  This can be seen as a dumb hole, i.e. an acoustic analogue of a black hole \cite{Unruh1981Experimental}. 
The presence in this new analogue system of effective horizons opens up new possibilities to explore the black hole evaporation with experiments  \cite{Weinfurtner2011Measurement}.

\subsection{Supersonic case (case 4 ) }

When the problem is supersonic everywhere 2 incoming waves are present upstream and 2 outgoing waves are present downstream (Fig. 5). Then there is no reflection and the scattering matrix is reduced to a $2 \times 2$ matrix that can be computed in the same way as previously, taking into account an evanescent mode on both sides of the discontinuity. The waves are mainly transmitted. Due to the characteristic of the negative energy wave, the transmission of the waves is always larger than 1.

\section{Scattering by a local increase  in the wall compliance \label{Sect5}}

\begin{figure}[h!]
\begin{center}
\includegraphics[width=0.7\columnwidth]{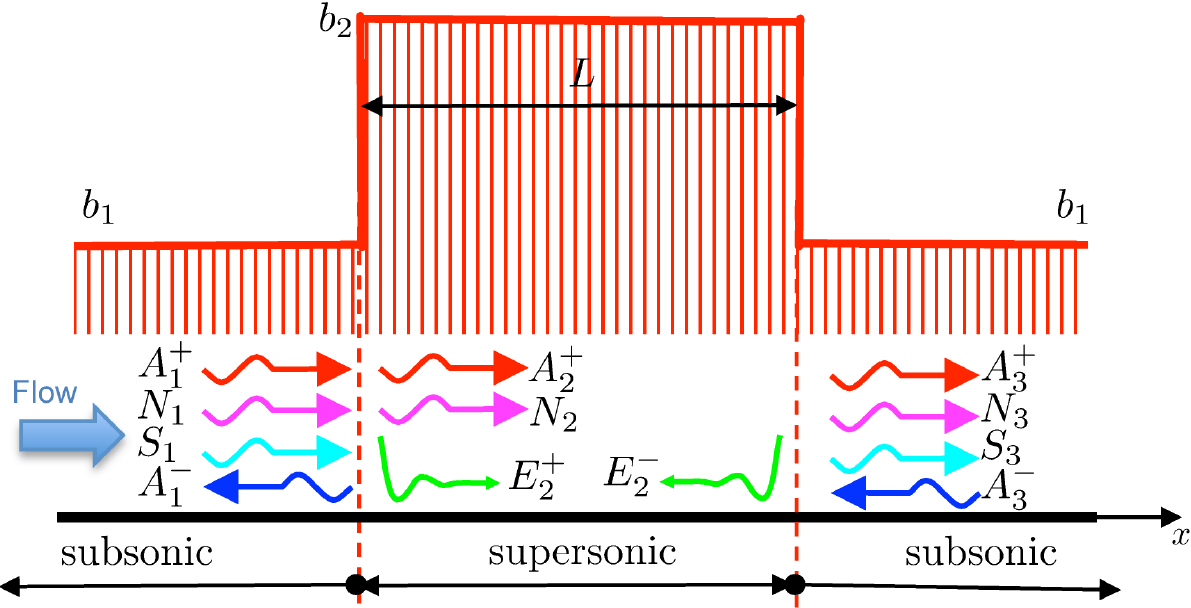}%

\begin{minipage}{10cm}
 {\small \textbf{Figure 10:} \textsl{Local increasing on in the wall compliance.  }}
 \end{minipage}
\end{center}
\end{figure}

A local increase in the wall compliance is depicted in Fig. 10. This configuration is computed as previously: the continuity of $\Vector{X}$ is applied at $x=0$ and at $x=L$, the propagation of the 2 modes $A_2^+$ and $N_2$ is taken into account between $x=0$ and $x=L$ and 2 decreasing evanescent modes are present on each side of the compliance bump  ($E_2^+$ and $E_2^-$). 
There are 8 unknowns $\Vector{C}= [a_3^+, \, n_3, \, s_3,\, a_1^-, \, a_2^+, \, n_2, \, e_2^+, \, e_2^-, \, ]^\mathbf T$ and 
four input values: $\Vector{A}= [a_1^+, \, n_1, \, s_1\, a_3^-]^\mathbf T$. From the 8 continuity relations, a vectorial relation can be written as $\Matrix{M_C}\Vector{C}=\Matrix{M_A}\Vector{A}$. The global $4 \times 4$ $\Matrix{S}$-matrix is composed of the first four lines of the matrix $\Matrix{M_C}^{-1} \Matrix{M_A}$. The absolute values of coefficients $ T_{A}^+=S_{a_3^+a_1^+}$,  $R_{A}^+=S_{a_1^- a_1^+}$,$T_{A}^-=S_{a_1^- a_3^-}$ and  $R_{A}^-=S_{a_3^+ a_3^-}$ are displayed in Fig. 11.

\begin{figure}[h!]
\begin{center}
\includegraphics[width=0.7\columnwidth]{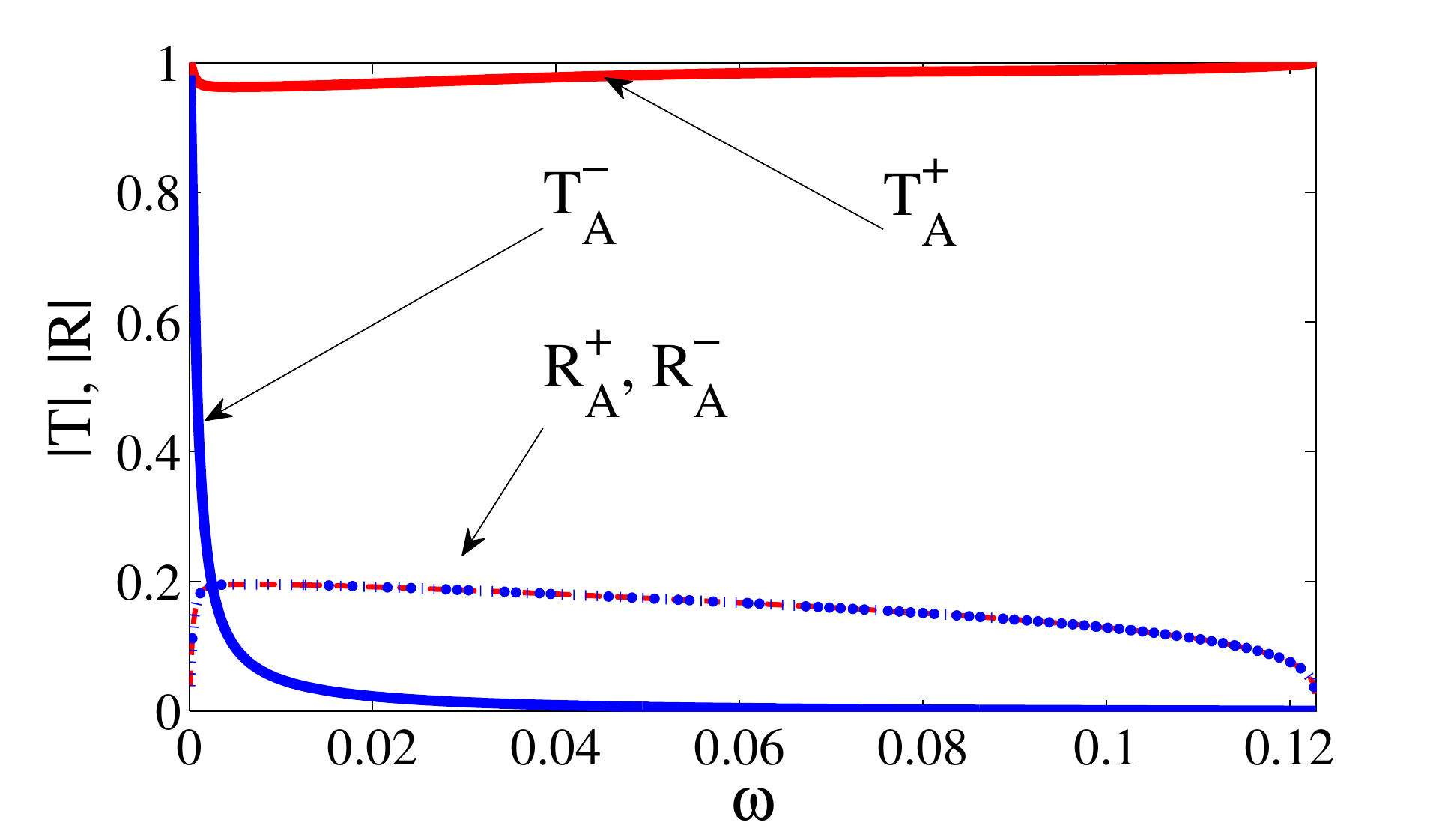}%

\begin{minipage}{10cm}
 {\small \textbf{Figure 11:} \textsl{(Color online) Absolute value of acoustical transmission and reflection coefficients in the flow direction $T_{A}^+$ and  $R_{A}^+$ and against the flow $T_{A}^-$ and  $R_{A}^-$  for $M=0.3$, $b_1=12$, $b_2=4$, $a_1=-a_2=4$ and $L=5$.}}
 \end{minipage}
\end{center}
\end{figure}

The acoustic transmission in the flow direction $ T_{A}^+$ is close to 1 while acoustic transmission against the flow $T_{A}^-$ is close to 0 because no wave can propagate against the flow. In this low frequency range, some acoustic is transmitted by the evanescent modes (tunneling effect). This system has been completely asymmetrised by the flow and act as an "acoustical diode" for a large range of frequencies.  

\section{Concluding remarks}

We have shown that the propagation of slow sound with flow at moderate Mach number have interesting and new properties. With such a system, it is possible to have subsonic and supersonic propagation and to make transition (soft or abrupt) from one regime to the other. The scattering properties of those transitions are very similar to what happens to light near a  white or a black hole. This analogy is useful in two ways. For instance, an acoustical analogue of a "black hole laser" \cite{Finazzi2010} can be studied as an inverse of the work done in Section 4 (Supersonic $\to$ Subsonic $\to$ Supersonic). On the other hand, this new acoustical analogy open opportunities to do simple experiments on these subjects.

\begin{acknowledgments} 
The authors wish to thank Renaud Parentani, Pierre Fromholz and Florent Michel for fruitful discussions on this subject and others. They also thank Gw\'ena\"el Gabard for his help on the temporal code used to produce Fig. 8.
 \end{acknowledgments}

\appendix

\section{Dispersion relation in $\Omega$ vs $k$} 
\label{AppendixI}

The dispersion curves are displayed on Fig. 12 in term of  $\Omega$ vs $k$. The thick continuous curves ($\Omega$-curves) represent the  Eq.  (\ref{eq:5}):  $\sqrt{k^2-\Omega^2} \tanh(\sqrt{k^2-\Omega^2}) -\tan (b \: \omega)\Omega^2/\omega  = 0$. At low frequencies, this curve depends only on $b$. The straight line represents $\Omega = \omega - k M$ and the solutions are found at the intersection of the $\Omega$-curves and of the straight line. 
In the displayed case, Eq.  (\ref{eq:5}) has 4 real solutions.  When $\omega$ increases at fixed $M$ (parallel translation of the straight line toward larger $\Omega$), two roots, labeled $S$ and $A^-$ on Fig. 12, become closer and closer. They coalesce for the frequency $\omega_{max}$. The same phenomenon occurs when  $M$ increases at fixed $\omega$ (rotation of the straight line toward larger negative slopes).

It can be also seen in Fig. 12 that if the slope of the straight line (given by $M$) is larger than the slope of the dispersion relation at the origin ($\mathrm{d}\Omega/ \mathrm{d}k = c_b$ where $c_b$ is the "slow sound" velocity given by Eq. (\ref{eq:5c})), represented by dashed lines in the figure, only two solutions can exist whatever $\omega$. In the 1D model, the slope at the origin is equal to $\pm 1/\sqrt{1- a_2 b/a_1}$. In order to have the same slope at the origin in the 1D and 2D model, we must have $a_2=-a_1$.

\begin{figure}[h!]
\begin{center}
\includegraphics[width=0.7\columnwidth]{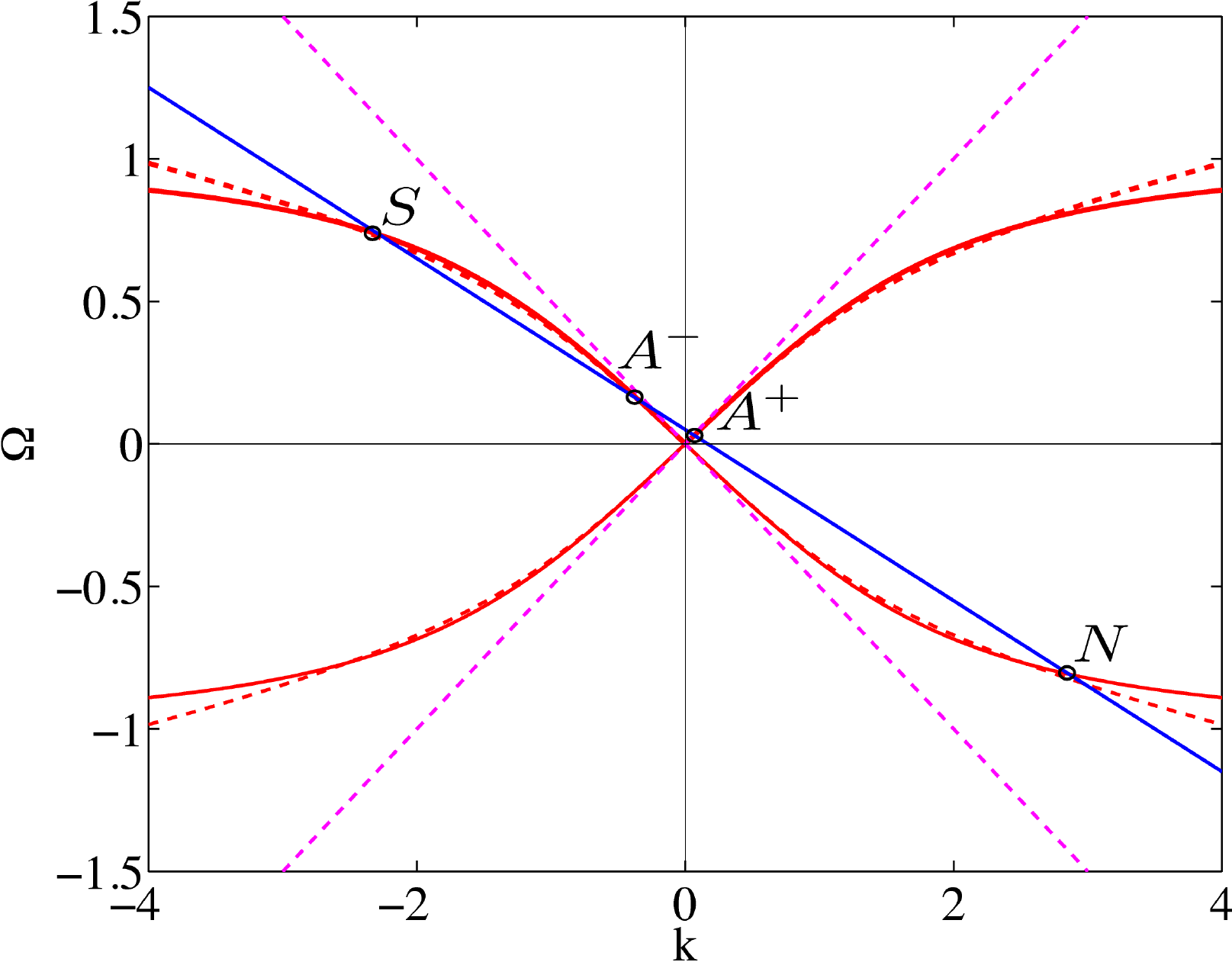}%

\begin{minipage}{10cm}
 {\small \textbf{Figure 12:} \textsl{(Color online) Solutions of the dispersion equation Eq.  (\ref{eq:5}) in term of  $\Omega$ versus $k$  for  $b=4$. 
The solution in term of of $\omega$ versus $k$ can be found at the intersections between these $\Omega$-curves and the straight line $\Omega = \omega - k M$ with $\omega=0.05$ and $M=0.3$. The thin dashed lines represent the slope at the origin and the thick dashed lines is the $\Omega$-curve of the 1D model (Eq. (\ref{eq:1D5}) with $a_2=-a_1=4$).}}
 \end{minipage}
\end{center}
\end{figure}

The $\Omega$ vs $k$ representation is also interesting because it allows the determination of the group velocity in the moving frame linked to the flow $c^M_g = \mathrm{d}\Omega/ \mathrm{d} k$. In our case, 3 solutions have a negative $c^M_g$ ($A^-,\:S,\:N$) and one has a positive $c^M_g$ ($A^+$). 
Depending on the sign of the curvature of the $\Omega$-curve, the group velocity at high $k$ can be larger or smaller than the group velocity at low $k$. These cases are usually referred to in the literature as "superluminal" and "subluminal" dispersion relations\cite{barcelo2011analogue}. The slow sound analogy has a "subluminal" dispersion relation.

\section{Modified unitary relation} 
\label{AppendixII}

When the wave is taken under the form  (\ref{eq:field}), the energy flux conservation (\ref{eq:1DEC}) between  the regions 1 and 2 can be written:
\begin{eqnarray}
 & &|a_1^+ |^2 - |n_1|^2 |+|s_1|^2 | -|a_1^+ |^2 =  \nonumber\\
 & &|a_2^+ |^2 - |n_2|^2 |+|s_2|^2 | -|a_2^+ |^2.
\end{eqnarray}
Splitting the incoming and the outgoing modes leads to:
\begin{eqnarray}
 & &|a_1^+ |^2 - |n_1|^2 |+|s_1|^2 | +|a_2^+ |^2 =  \nonumber\\
 & &|a_2^+ |^2 - |n_2|^2 |+|s_2|^2 | +|a_1^+ |^2.
\end{eqnarray}
which can be written:
\begin{equation}
\overline{\Vector{A}}^T \Matrix{J} \Vector{A} = \overline{\Vector{B}}^T \Matrix{J} \Vector{B} 
\end{equation}
where the vectors $\Vector{A}$ and $\Vector{B}$ are given in Eq. (\ref{eq:Scatt}) and $\Matrix{J} = \mathrm{diag}(1,\, -1,\, 1,\, 1)$.
Using the definition of the scattering matrix, it can be written: 
 \begin{equation}
 \overline{\Vector{A}}^T\Matrix{J}\Vector{A}=\overline{\Vector{A}}^T\overline{ \Matrix{S}}^T \Matrix{J} \Matrix{S}\Vector{A}
\end{equation}
This relation, valid whatever $\Vector{A}$, leads to the Eq. (\ref{eq:MatConserv}).

\end{document}